\documentclass[a4paper,11pt]{article}
\usepackage{jinstpub}
\usepackage{flafter}

\newcommand{\xtcptitle}{XTCP: an open-source FPGA TCP endpoint for readout over 10 and 25 Gigabit Ethernet}
\title{\xtcptitle}
\hypersetup{pdftitle={\xtcptitle},pdfauthor={Tao Xue},
  pdfsubject={FPGA implementation and single-connection measurements of a TCP readout endpoint}}
\author[a,b,1]{Tao Xue\note{Corresponding author.}}
\affiliation[a]{Key Laboratory of Particle \& Radiation Imaging (Tsinghua University), Ministry of Education,\\
Qinghuayuan Street No.~1, Beijing, China}
\affiliation[b]{Department of Engineering Physics, Tsinghua University,\\
Qinghuayuan Street No.~1, Beijing, China}
\emailAdd{xuetaothu@tsinghua.edu.cn}
\abstract{XTCP is an open-source implementation of the Transmission Control
Protocol (TCP) on field-programmable gate arrays (FPGAs) for high-rate detector
readout. It targets sustained single-connection throughput near the TCP payload
limit. In separate 10 Gigabit Ethernet transfers on the same Kintex UltraScale+ KU5P card and Linux host, XTCP and
SiTCP-XG each delivered a median application-level throughput of 9.49~Gbit/s
towards the host. Towards the FPGA, the medians were 9.49~Gbit/s for XTCP and
5.76~Gbit/s for the tested SiTCP-XG configuration. The XTCP prototype used more lookup
tables and on-chip memory. In a 96~GiB echo test spanning repeated
sequence-number wraps, recording when an acknowledgement first crossed the recovery
boundary allowed four fast recoveries that the earlier signed comparison would
have blocked. On one 25 Gigabit Ethernet bitstream, increasing the IP maximum
transmission unit from 1500 to 9000~bytes raised median one-way throughput
from 23.30 to 24.70~Gbit/s towards the host and from 22.09 to 24.08~Gbit/s
towards the FPGA. At 9000~bytes these rates reached 99.7\% and 97.1\% of the TCP payload
limit, respectively. In a separate echo test, the host verified all 8.77~TB
returned by a jumbo-frame prototype over one connection in 3595~s, averaging
19.51~Gbit/s per direction.}

\keywords{Front-end electronics for detector readout; Data acquisition concepts; Digital electronic circuits}
\newcommand{\gbs}{\ensuremath{\mathrm{Gbit/s}}}

\begin{document}
\maketitle

\section{Introduction}
\label{sec:intro}
In a continuous detector readout system, a front end must buffer data when the
host temporarily receives them more slowly than the detector produces them. Triggerless readout for CDEX-50
is one example~\cite{yang2024cdex}. Ethernet links have also been used to
transfer pixel-detector data from a field-programmable gate array (FPGA) to a
host~\cite{li2020}. The Transmission Control Protocol (TCP) provides an ordered
byte stream through a standard socket. An FPGA implementation must track which
bytes the peer has acknowledged and how much receive buffer space is available.

Uchida demonstrated Gigabit Ethernet readout with the SiTCP hardware
processor~\cite{uchida2008}. SiTCP-XG extended the approach to 10 Gigabit
Ethernet~\cite{sitcpxg}. For the CMS data acquisition system, Bauer et
al.\ implemented a unidirectional 10 Gigabit Ethernet TCP sender for Linux
receivers~\cite{bauer2014}. Sidler et al.\ addressed a different requirement,
the number of simultaneous TCP connections on an FPGA~\cite{sidler2015}.
The open-source Limago stack extended this architecture to 100 Gigabit
Ethernet~\cite{ruiz2019}.

XTCP (eXtreme TCP) is an open-source endpoint for transferring a continuous
stream between FPGA readout logic and a standard host socket. Its synthesizable
source is packaged as a configurable Vivado IP core. The three build configurations support
10 Gigabit Ethernet (10GbE), 25 Gigabit Ethernet (25GbE) with standard frames,
and 25GbE with jumbo-frame support. Each endpoint serves one connection.
This study evaluates application-level throughput, the number of TCP payload
bits delivered per unit time. Separate one-way transfers characterize the
transmit and receive paths. Two IP maximum transmission units (MTUs) were
tested on one 25GbE bitstream to measure the effect of frame size at fixed
hardware resources. Sustained echo tests examined data integrity across
repeated wraps of the 32-bit TCP sequence space.

\section{Endpoint architecture}
\label{sec:design}

\subsection{Transmit and receive paths}
\begin{figure}[htbp]
\centering
\includegraphics[width=.97\linewidth]{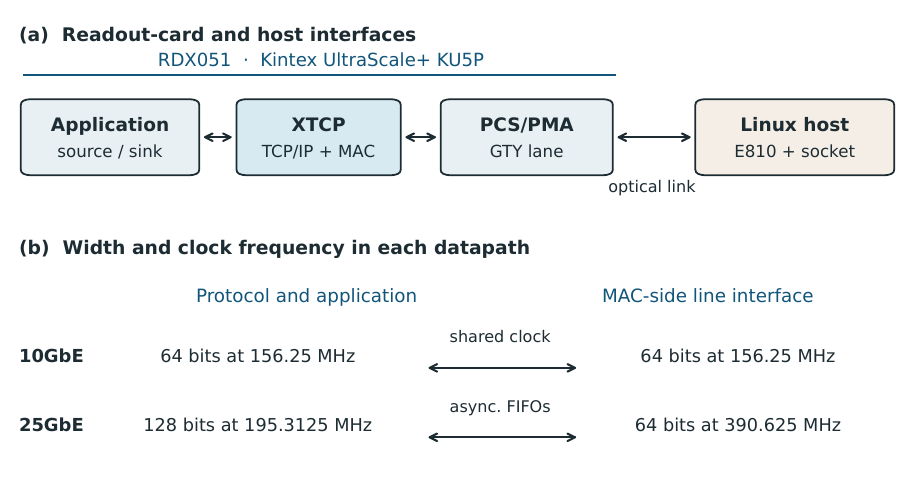}
\caption{Readout path for the v1.0.0 measurements on the RDX051 and clock
frequencies for the two link speeds. Arrows indicate bidirectional interfaces. At 25 Gigabit Ethernet separate transmit
and receive asynchronous first-in, first-out buffers (FIFOs) connect the two
clock domains. The differential transceiver reference is 156.25~MHz at both
link rates.}
\label{fig:system}
\end{figure}

The RDX051 readout card connects the FPGA application to a host socket over
one optical lane (figure~\ref{fig:system}). The application produces or
consumes a stream of bytes. XTCP implements the TCP connection state, packet
processing and Ethernet media access control (MAC) in programmable logic.
A BASE-R physical coding sublayer (PCS) and physical medium attachment (PMA)
connect the MAC to the serial transceiver. The host opens a TCP connection to
the configured IPv4 address and port. A separate User Datagram Protocol (UDP)
register interface on the same link provides access to configuration registers
and counters, including the number of bytes consumed by the FPGA application.

The transmit interface accepts a data word and a count of its valid bytes,
allowing a transfer to end on a partial word. The endpoint asserts an
almost-full signal to apply backpressure to the source. It stores accepted
bytes in a circular buffer until the peer acknowledges them. The packet
builder selects a range of bytes subject to the advertised receive window,
the available send credit and the negotiated maximum segment size (MSS).
It constructs the headers, assembles the payload and calculates the checksum.
Two frame buffers allow one frame to be assembled while the preceding frame
is transmitted. A validated cumulative acknowledgement (ACK) releases the
corresponding bytes for reuse. Until then, the sender can retransmit them
while accepting new application data.

On reception, an input queue buffers frames for parsing. The endpoint
checks frame and packet lengths, checksums, connection identifiers and TCP
sequence numbers before delivering data to the application. It writes
contiguous payload bytes to the application's circular buffer using a byte
address and per-byte write enables. A separate reorder buffer can retain
segments received beyond a gap. The application advances its read pointer
as it consumes data. The advertised receive window depends on the free space
in the application buffer, a configurable limit and the negotiated window scale.

\begin{figure}[htbp]
\centering
\includegraphics[width=.93\linewidth]{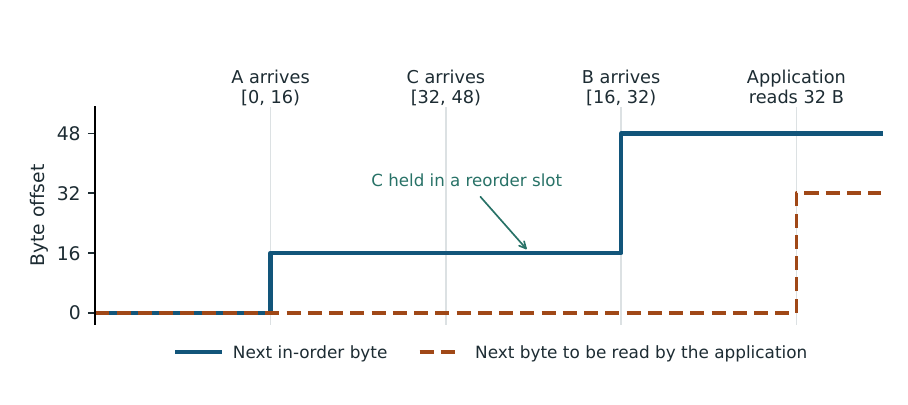}
\caption{Illustrative receive sequence with byte offsets relative to the first
payload byte. Segments arrive in the order A, C and B. After B fills the gap and
the endpoint commits C, the cumulative ACK advances to offset 48, acknowledging
bytes through offset 47. The application subsequently consumes 32~B. The horizontal spacing
indicates event order. ACK transmission and internal memory cycles are omitted.}
\label{fig:receive}
\end{figure}

Figure~\ref{fig:receive} illustrates the distinction between acknowledging
data and releasing receive buffer space. Filling a gap can advance the
cumulative ACK while the application read pointer remains stationary.
Conversely, consuming data can increase the advertised window while the
next expected sequence number remains unchanged~\cite{rfc9293}.

\subsection{Datapath width, clocks and on-chip storage}
The 10GbE implementation transfers eight bytes per clock cycle at
156.25~MHz. Both 25GbE configurations use a 128-bit application interface
at 195.3125~MHz and a 64-bit MAC-side interface at 390.625~MHz. Each can
transfer 25~\gbs{} before protocol overhead. Doubling the datapath width
gives the protocol logic a 5.12~ns clock period, while the line-side
datapath operates with a period of 2.56~ns. Byte enables preserve partial
words across the width conversion. The design uses Ethernet framing and
cyclic redundancy check (CRC) logic from \emph{verilog-ethernet}~\cite{forencichEthernet},
width conversion and asynchronous FIFOs from \emph{verilog-axis}~\cite{forencichAxis},
and the PCS/PMA from version 5.0 of the AMD 10G/25G Ethernet
subsystem~\cite{amdPG210}.

Table~\ref{tab:resources} lists the resources used after place and route in
Vivado 2025.2 for the three XTCP v1.0.0 configurations on a Kintex UltraScale+
KU5P FPGA (\texttt{xcku5p-ffvb676-2-i}). The endpoint counts include the
MAC and frame processing logic. The example application has its own receive
buffer. All configurations allocate 256~KiB to the transmit buffer for
unacknowledged data. The 10GbE design allocates 64~KiB to the input frame
buffer and 256~KiB to the application receive buffer. Each buffer is
512~KiB in the 25GbE designs. All TCP buffering is on chip. At a readout
rate of 24.70~\gbs{}, an initially empty 256~KiB transmit buffer holds about
85~$\mu$s of input. Bytes awaiting acknowledgement and the almost-full
threshold reduce the capacity available during a host service interruption.

\begin{table}[htbp]
\centering
\caption{FPGA resource utilization after routing for XTCP v1.0.0. The endpoint
columns count lookup tables (LUTs), flip-flops (FFs), block RAM primitives,
UltraRAM blocks and digital signal processing (DSP) blocks. The board columns
include the application, PCS/PMA and instrumentation. WNS is the worst setup
slack reported by the timing analysis.}
\label{tab:resources}
\small
\setlength{\tabcolsep}{4pt}
\begin{tabular}{lrrrrrr|rr}
\hline
& \multicolumn{6}{c|}{Endpoint} & \multicolumn{2}{c}{Board design}\\
Configuration & LUTs & FFs & RAMB36 & RAMB18 & URAM & DSP & LUTs & WNS (ns)\\
\hline
10GbE standard & 14\,156 & 8\,807 & 82 & 0 & 4 & 1 & 17\,095 & 0.043\\
25GbE standard & 23\,215 & 11\,272 & 198 & 2 & 4 & 1 & 26\,740 & 0.127\\
25GbE jumbo    & 23\,385 & 11\,227 & 216 & 2 & 32 & 1 & 26\,851 & 0.032\\
\hline
\end{tabular}
\end{table}

Most of the additional memory in the 25GbE jumbo-frame design is allocated
to receive-side reordering. Both 25GbE designs provide 64 reorder slots.
Each slot occupies 2~KiB in the standard-frame design and 16~KiB in the
jumbo-frame design to accommodate a TCP payload of 8960~B. The eightfold
increase in slot capacity raises the UltraRAM count from four to 32. Larger
frame buffers and clock-domain-crossing FIFOs require 18 additional RAMB36
blocks. The endpoint logic increases by 170 LUTs.
The jumbo-capable implementation also accepts a 1500-byte MTU, allowing both
frame sizes to be tested with a single FPGA configuration.

\section{Measurement methods}
\label{sec:methods}

\subsection{Experimental setup and transfer tests}
The three XTCP v1.0.0 configurations were programmed successively into the same
RDX051 card. A single optical lane connected the card directly to an Intel E810
network interface controller (NIC) in a Linux host (Core Ultra 9 285HX,
kernel 6.17.0-14-generic). The host initiated all TCP connections. The link used
IPv4 without VLAN tagging. The negotiated link rate and disabled forward error
correction (FEC) setting were verified at the port. Each link rate used an optical
cable rated for that speed. In a preliminary 10GbE test with the 25GbE cable,
the host recorded 63 receive CRC errors and the FPGA recorded 17 frame check
sequence errors. With the corresponding 10GbE and 25GbE cables, neither
counter increased during the throughput measurements.

Four settings were tested. The standard-frame 10GbE and 25GbE designs used an
IP MTU of 1500~B. The jumbo-capable 25GbE design was tested at both 1500 and
9000~B. At each setting, 11 echo tests covered payload lengths from 1 to
65\,539~B, including 7 and 8, 15 through 17, 1460 and 1461, and 8960 and
8961~B. Transfers of 1\,048\,593~B in each direction checked the application
byte counters. Together these 52 functional checks covered partial words and
transfers spanning several TCP segments.

For each setting, the host ran three workloads on separate connections.
Each workload was repeated three times, giving 36 timed transfers. For
FPGA-to-host transfers, the FPGA generated a deterministic byte pattern and
a C client verified each received byte. For host-to-FPGA transfers, the
client sent the pattern and waited for the acknowledged-byte count to reach
the requested value. The FPGA application byte counter provided an independent
check on the delivered length. Each one-way transfer comprised $4~\mathrm{GiB}+17~\mathrm{B}$.
For echo, the host transmitted and received concurrently on one connection.
The FPGA read the received bytes from the application buffer and returned
them over TCP. The host verified all returned bytes and waited for the
outstanding transmit bytes to be acknowledged. Each echo transfer comprised
$1~\mathrm{GiB}+17~\mathrm{B}$ per direction. The extra 17~B exercised a
partial final word at both interface widths.

The client measured elapsed time with a monotonic clock, starting after
connection setup and ending at payload completion. For host-originated transfers, this
interval included the wait for the final acknowledgement. FPGA application
byte counters were read through the UDP interface. The host used its default
socket buffer sizes and adaptive interrupt moderation for transmit and receive.
Client CPU utilization was calculated as the sum of user and system CPU time
divided by elapsed time, with 100\% corresponding to one logical core. Changes
in the operating system's CPU counters gave the mean number of busy logical
cores across the host, including interrupt handling and other host activity.

\subsection{Application-level throughput and theoretical limit}
The application-level throughput in one direction is $8N/T$ in bit/s, where
$N$ is the number of TCP payload bytes delivered to the receiving application
during the interval $T$ in seconds. Each byte-stream position is counted once.
Protocol headers and retransmissions contribute no additional payload bytes. For echo, $N$ is the number of returned bytes verified by the
host. The reported echo throughput is therefore per direction. Throughput is
expressed in decimal units. Transfer sizes expressed in GiB use
$1~\mathrm{GiB}=2^{30}~\mathrm{B}$.

For a one-way stream of full-sized segments, the theoretical maximum TCP
payload throughput is
\begin{equation}
R_{\rm max}=R_{\rm MAC}\,\frac{M-40\,\mathrm{B}}{M+38\,\mathrm{B}},
\label{eq:ceiling}
\end{equation}
where $R_{\rm MAC}$ is the nominal Ethernet MAC rate and $M$ is the IP MTU.
The IPv4 and TCP headers each occupy 20~B in these tests. The denominator
includes the 14-byte Ethernet header, the 4-byte frame check sequence,
8 byte times for the preamble and start-of-frame delimiter, and 12 byte
times for the interframe gap~\cite{ieee8023}. The model assumes that the
reverse channel can carry the ACK traffic. TCP timestamps were not negotiated on any
of the 36 connections. Linux reported send MSS values of 1460
and 8960~B for MTUs of 1500 and 9000~B, respectively. At an MTU of 1500~B,
eq.~\eqref{eq:ceiling} gives 9.493~\gbs{} at 10GbE and 23.732~\gbs{} at
25GbE. At an MTU of 9000~B, the 25GbE limit is 24.784~\gbs{}. During echo,
ACKs share each link direction with data segments, and the application
couples transmission to reception.

\subsection{Sustained echo and integrity verification}
The sustained tests used three prototype bitstreams, one for each configuration,
before the v1.0.0 builds. A sender generated payload bytes from their positions
and a seed. A receiver checked every byte returned on the same connection.
The tests were scheduled for one hour, allowing approximately 5~s for outstanding
data to drain and the connection to close. Each run produced 359 complete,
non-overlapping 10~s intervals. The overall mean includes all bytes returned
through the end of the transfer. For the 10GbE prototype, verified receive-byte
counts were interpolated at the interval boundaries. For the two 25GbE prototypes,
the interval throughput was the smaller of the sent and verified receive rates
in that interval. The fifth percentile was taken as the 18th of the 359 interval
rates sorted in ascending order.

The 10GbE prototype used an Intel X710 host adapter. Both 25GbE prototypes
used the E810 host with adaptive interrupt moderation, a 480~KiB limit on
the FPGA receive window and the same pair of logical cores for the host
receive and transmit threads. All sustained runs used the same byte-verification
criterion. The fixed-length v1.0.0 measurements used the E810 host for all
three configurations. The supplementary data identify each programmed image
by its bitstream hash.

\section{Results}
\label{sec:results}

\subsection{Single-connection throughput and host CPU utilization}
\begin{figure}[htbp]
\centering
\includegraphics[width=\linewidth]{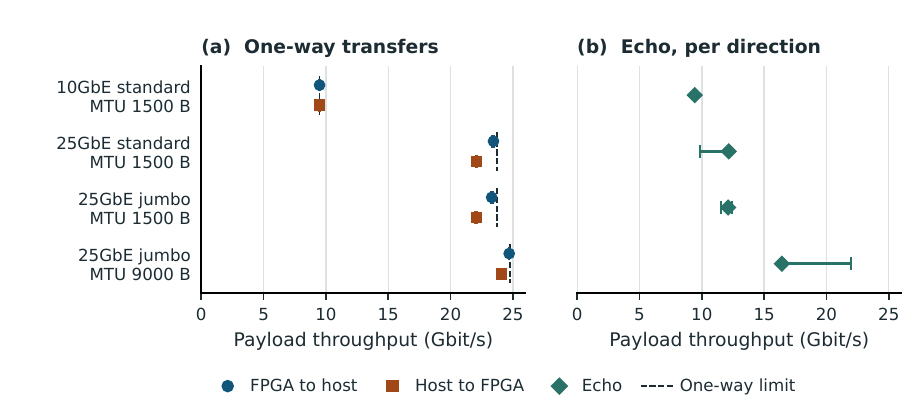}
\caption{Fixed-length transfers over one TCP connection at a time. Symbols
are medians, and horizontal bars span the minimum and maximum of three
trials. The dashed marks in (a) indicate the theoretical limits from
eq.~\eqref{eq:ceiling}. Each point in (b) counts the echoed payload once
per direction.}
\label{fig:throughput}
\end{figure}

Table~\ref{tab:rates} gives the median throughput for each v1.0.0 configuration.
Figure~\ref{fig:throughput} shows the minimum and maximum of the three trials
at each setting. For standard frames at 10GbE, both one-way medians were
within 0.1\% of the theoretical TCP payload limit. For the standard-frame
25GbE design, the FPGA-to-host and host-to-FPGA medians reached 98.8\% and
93.0\% of that limit, respectively. Neither endpoint recorded a TCP
retransmission in these one-way tests.

\begin{table}[htbp]
\centering
\caption{Application-level throughput for the XTCP v1.0.0 configurations.
Measured rates are medians of three connections per workload. All rates are in
Gbit/s. The theoretical limit is for one-way transfer of full-sized segments.}
\label{tab:rates}
\small
\setlength{\tabcolsep}{5.2pt}
\begin{tabular}{lrrrrr}
\hline
Configuration & MTU (B) & Limit & FPGA to host & Host to FPGA & Echo per direction\\
\hline
10GbE standard & 1500 & 9.493 & 9.487 & 9.486 & 9.427\\
25GbE standard & 1500 & 23.732 & 23.443 & 22.068 & 12.155\\
25GbE jumbo & 1500 & 23.732 & 23.296 & 22.093 & 12.115\\
25GbE jumbo & 9000 & 24.784 & 24.702 & 24.077 & 16.419\\
\hline
\end{tabular}
\end{table}

On the same jumbo-capable bitstream, increasing the MTU from 1500 to 9000~B
raised the median throughput from 23.296 to 24.702~\gbs{} from FPGA to host
and from 22.093 to 24.077~\gbs{} from host to FPGA. At the larger MTU these
values correspond to 99.67\% and 97.15\% of the theoretical TCP payload limit.
At fixed payload throughput, increasing the MSS from 1460 to 8960~B reduces
the number of full-sized segments per second by a factor of 6.14. In the
FPGA-to-host tests, client CPU utilization, including byte verification,
decreased from 67.1\% to 48.5\% of one logical core. Mean CPU utilization
across the host fell from 1.17 to 0.70 busy cores. In the reverse direction,
client utilization decreased from 44.0\% to 38.2\%, and the corresponding
host-wide values were 0.57 and 0.58 busy cores.

In the echo tests, the FPGA application returned received data while the
host transmit and receive threads ran concurrently. At 10GbE the median
echo throughput was 9.427~\gbs{} per direction. For the jumbo-capable 25GbE
bitstream, the median was 12.115~\gbs{} at an MTU of 1500~B and
16.419~\gbs{} at 9000~B. At 9000~B the three runs ranged from 16.346 to
21.942~\gbs{}, with zero TCP retransmissions recorded at either endpoint.
The standard-frame 25GbE bitstream at a 1500-byte MTU gave a median of
12.155~\gbs{} and a minimum of 9.852~\gbs{}.

\subsection{Sensitivity to host interrupt moderation}
In a separate test of the standard-frame 25GbE prototype, each one-way
transfer carried 16~GiB. In each of two FPGA programming cycles, three
connections per direction were measured with adaptive receive interrupt
moderation and three with a fixed receive interval of 8~$\mu$s. The host
source ports were matched across settings, and transmit interrupt moderation
remained adaptive. For host-to-FPGA transfers, the median throughput in the
two cycles was 23.023 and 23.022~\gbs{} with the adaptive setting, and
23.222 and 23.407~\gbs{} at 8~$\mu$s. In the reverse direction the medians
were 23.406 and 23.325~\gbs{} with the adaptive setting, and 23.496 and
23.478~\gbs{} at 8~$\mu$s. The fixed interval gave a higher median in
both directions in both cycles. In the host-to-FPGA direction, the NIC also
processes the returning ACKs. The bitstream, transfer length and timing
procedure differed from those used in table~\ref{tab:rates}.

\subsection{Retransmission recovery and sequence-number wraparound}
\begin{figure}[htbp]
\centering
\includegraphics[width=.97\linewidth]{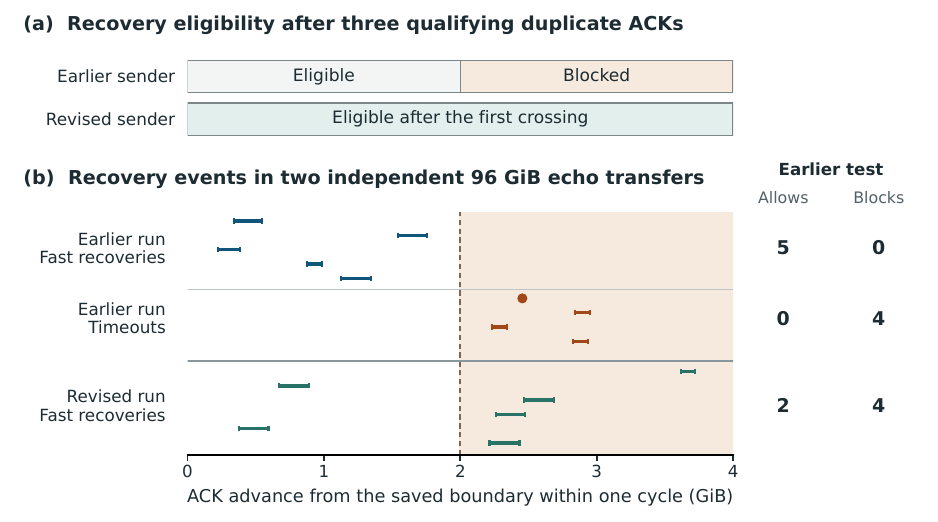}
\caption{Fast recovery across a 4~GiB sequence cycle. (a) Boundary tests
after the first ACK crossing. (b) All events in two independent, unpaired
runs, in time order within each group. Bars bound the ACK advance from the
saved boundary modulo 4~GiB. A dot denotes coincident bounds. Shading and
counts apply the earlier comparison to both runs.}
\label{fig:recovery}
\end{figure}

To avoid repeated fast retransmits for the same outstanding data, the
10GbE sender checks whether the cumulative ACK has passed a saved recovery
boundary~\cite{rfc6582}. The 32-bit TCP sequence number wraps every 4~GiB. Once
the ACK has advanced more than 2~GiB, a signed comparison can place it behind
a boundary it has already crossed. At 9.49~\gbs{} this half-cycle takes
1.81~s. The revised sender records the first validated ACK strictly beyond
the boundary and clears this flag when it sets a new boundary
(figure~\ref{fig:recovery}(a)).

FPGA counters sampled about every 0.10~s bracketed 15 events in two
96~GiB echo runs (figure~\ref{fig:recovery}(b)). On the earlier connection,
all five fast recoveries fell in the allowed half-cycle and all four
timeouts in the blocked half-cycle. Three qualifying duplicate ACKs
preceded each timeout. The stale comparison blocked fast retransmit,
leaving the sender to wait for its timer. The revised sender entered
fast recovery six times with zero timeouts. At these six sequence positions
the earlier comparison would have allowed two entries and blocked four.

In separate sustained
echo runs, the earlier and revised designs recorded 121 and five timeouts,
respectively. The number of 10~s intervals below 9~\gbs{} fell from 113 to
five out of 359. In a separate 1~GiB transfer the revised design recorded
one timeout and returned all bytes at 4.478~\gbs{}. The transfer covered less
than half the sequence space, so its completion time alone cannot be used
to diagnose sequence-number wraparound.

Controlled tests of the 10GbE prototype covered packet loss, duplication,
reordering, a zero advertised window and reconnection. During one 512~MiB
transfer, the receive-side impairment dropped 180 packets. The FPGA recorded
one retransmission timeout and replayed 180 transmit descriptors. The host verified
the entire payload in 1.458~s. The 25GbE designs were tested for receive-window
reopening, segment-length boundaries and sustained data integrity. For the
jumbo-frame design, simulations covered unaligned payload lengths and
independently verified the Ethernet CRC and the IPv4 and TCP checksums of
54 test frames at length boundaries.

\subsection{Sustained echo throughput}
In each of the three sustained tests, the FPGA returned the entire transmitted
stream and the host verified every byte (table~\ref{tab:long}). The 25GbE jumbo-frame
prototype returned 8\,768\,350\,064\,640~B over one connection in
3595.002~s, corresponding to more than 2000 complete TCP sequence-number
cycles in each direction. Its mean echo throughput was 19.512~\gbs{} per
direction, compared with 13.283~\gbs{} for the standard-frame prototype.
The two runs used the same host and receive-window limit. Both the MTU
and the FPGA buffer allocations differed.

\begin{table}[htbp]
\centering
\caption{Continuous echo on three engineering prototypes, one connection per
row. Returned data are decimal TB, using $10^{12}$~B per TB. The three
rightmost columns give throughput per direction in Gbit/s. The percentile
and minimum use the complete 10~s windows.}
\label{tab:long}
\small
\setlength{\tabcolsep}{6pt}
\begin{tabular}{lrrrrr}
\hline
Prototype & Elapsed (s) & Returned (TB) & Mean & 5th percentile & Minimum\\
\hline
10GbE standard & 3595.002 & 4.246 & 9.450 & 9.457 & 8.503\\
25GbE standard & 3595.001 & 5.969 & 13.283 & 11.395 & 11.306\\
25GbE jumbo    & 3595.002 & 8.768 & 19.512 & 16.447 & 16.439\\
\hline
\end{tabular}
\end{table}

\begin{figure}[htbp]
\centering
\includegraphics[width=.98\linewidth]{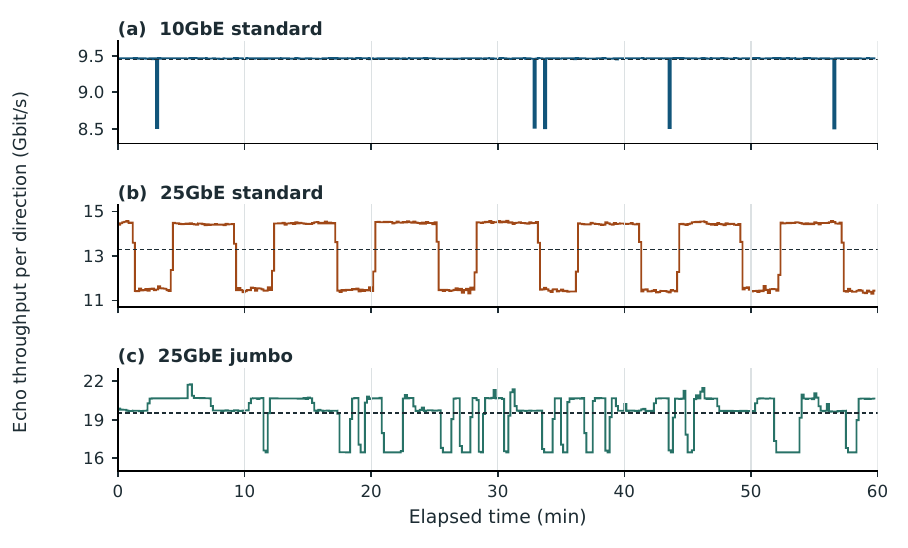}
\caption{Measured echo throughput per direction in 359 consecutive 10~s intervals for each
prototype. Dashed lines are whole-run means including the final partial
interval. Panel (a) uses the verified receive-byte count. Panels (b) and
(c) use the smaller of sent and verified receive bytes in each interval.
The vertical scales differ to resolve variation within each configuration.}
\label{fig:sustained}
\end{figure}

Figure~\ref{fig:sustained} gives the time dependence of the echo throughput.

A few low-throughput intervals in the 10GbE run lowered the overall mean
below the fifth percentile of the interval rates. For the standard-frame
25GbE prototype, the interval throughput varied repeatedly between about
11.4 and 14.5~\gbs{}. For the jumbo-frame prototype, 71 of 359 intervals
fell below 18~\gbs{}, including a continuous period of 120~s. Neither
endpoint recorded a TCP retransmission during that run. The duration of
the lower-throughput intervals is relevant to buffering a continuous
input stream. Determining their cause would require time-resolved
measurements of queue occupancy and service at both endpoints.

\section{Comparison with other FPGA TCP implementations}
\label{sec:comparison}

\subsection{Single-connection measurements on a common 10GbE platform}
SiTCP-XG~\cite{sitcpxg} and the 10GbE XTCP prototype (table~\ref{tab:long})
used the same KU5P card, optical link and Linux host with an Intel X710 NIC
(MTU 1500~B, MSS 1460~B). The same client timed six 32~GiB transfers in two
groups of three for each endpoint and direction (table~\ref{tab:matched}).
Timing ended when the host verified the final byte for FPGA-to-host transfers
or confirmed the FPGA application byte count for host-to-FPGA transfers.
Table~\ref{tab:rates} reports shorter transfers on the E810 host.

\begin{table}[htbp]
\centering
\caption{Single-connection medians of six transfers on the common 10GbE platform.
Client CPU time includes user and system time per GiB of payload.}
\label{tab:matched}
\small
\setlength{\tabcolsep}{4pt}
\begin{tabular}{lrrrr}
\hline
& \multicolumn{2}{c}{Throughput (Gbit/s)} &
  \multicolumn{2}{c}{Process CPU time (s/GiB)}\\
Endpoint & FPGA$\to$host & Host$\to$FPGA & FPGA$\to$host & Host$\to$FPGA\\
\hline
SiTCP-XG & 9.491 & 5.758 & 0.322 & 0.165\\
XTCP prototype & 9.493 & 9.488 & 0.318 & 0.064\\
\hline
\end{tabular}
\end{table}

Both endpoints delivered 9.49~\gbs{} to the host at similar client CPU cost.
In the reverse direction, SiTCP-XG (128~KiB transmit and 32~KiB application
receive buffers) and XTCP (256~KiB application receive buffer) advertised
median final windows of 32\,752 and 262\,124~B, respectively. Linux classified
99.998--100\% of sender-busy time as receive-window limited for SiTCP-XG
and 99.758--99.962\% for XTCP. This time fraction was near unity even at 9.488~\gbs{}. The 1.65-fold throughput difference
applies to these configurations. Isolating the contribution of window size
would require varying it independently.

A separate SiTCP-XG echo test stopped at a payload mismatch after 1.425~s.
The X710 receive CRC error count rose by 19. The counters do not identify the
cause of either error, and this run provides no sustained throughput estimate.

\subsection{Resource utilization and published throughput measurements}
Table~\ref{tab:peerresources} specifies the modules counted for each endpoint
or TCP offload engine (TOE). The two 10GbE endpoint counts include the MAC.
The complete SiTCP-XG and XTCP prototype designs, including the application,
PCS/PMA and debug logic, used 8423 and 17\,023 LUTs, respectively. Their
application receive buffers occupied eight and 64 block RAM tiles. XTCP also
allocated on-chip memory to transmit retention, input frames and segment
reordering.

\begin{table}[htbp]
\centering
\caption{FPGA resource utilization for the modules listed. One block RAM
tile represents one RAMB36 primitive or two RAMB18 primitives. The first four rows give post-route counts on the KU5P used in this study.
The remaining rows give the counts reported in the cited sources. A dash denotes an
unreported UltraRAM count, and n/a denotes a device without UltraRAM.}
\label{tab:peerresources}
\small
\setlength{\tabcolsep}{3.5pt}
\begin{tabular}{llrrrr}
\hline
Design and module & FPGA & LUTs & FFs & BRAM tiles & URAM\\
\hline
SiTCP-XG, endpoint & KU5P & 5406 & 9954 & 70.5 & 0\\
XTCP 10GbE prototype, endpoint & KU5P & 13\,802 & 8787 & 82 & 4\\
XTCP 25GbE standard, endpoint & KU5P & 23\,215 & 11\,272 & 199 & 4\\
XTCP 25GbE jumbo, endpoint & KU5P & 23\,385 & 11\,227 & 217 & 32\\
\hline
TOE25G-IP, TOE~\cite{dg25core} & KU5P & 3904 & 3869 & 20 & 2\\
Sidler et al., TCP/IP stack~\cite{sidler2015} & VX690T & 19\,026 & 20\,611 & 279 & n/a\\
Limago, TOE~\cite{ruiz2019} & VU9P & 36\,469 & 36\,229 & 247.5 & --\\
\hline
\end{tabular}
\end{table}

Design Gateway's single-connection TOE25G-IP has a 128-bit user port with
a minimum clock frequency of 195.3125~MHz. Its KU5P example excludes the
MAC and allocates 64~KiB to transmit data, 16~KiB to transmit packets and
64~KiB to receive data~\cite{dg25core}. Its PC demonstration reported
22.63~\gbs{} from FPGA to host with 8960-byte payloads and host-side pattern
checking, and 21.24~\gbs{} from host to FPGA with FPGA-side checking.
Each value is for one 68.72~GB transfer~\cite{dg25demo}. The XTCP jumbo-frame
medians were 24.70 and 24.08~\gbs{}, respectively. The host, transfer length
and verification procedure differed. The vendor describes the resource
example and the throughput demonstration separately.

At an MSS of 536~B, Sidler et al.\ measured aggregate application throughput
of 8.5~\gbs{} into the FPGA over 25 connections and 8.6~\gbs{} from the FPGA
over 500 connections. Their stack resource count excludes the network and
external-memory interfaces~\cite{sidler2015}. The Limago TOE entry is for
10\,000 supported connections with window scaling disabled. Its published
single-connection rates include Ethernet, IP and TCP headers. Its host
application measurements used up to 30 concurrent connections~\cite{ruiz2019}.
The XTCP rates in table~\ref{tab:rates} count payload on one connection.

\section{Conclusions}
The open-source XTCP endpoint sustains high application-level throughput on
one connection. At 10GbE, XTCP and SiTCP-XG both delivered 9.49~\gbs{} to
the host. XTCP delivered 1.65 times the host-to-FPGA throughput of SiTCP-XG
in the configurations tested, which differed in receive buffer size and
advertised window. The XTCP prototype used 13\,802 LUTs and four UltraRAM
blocks, compared with 5406 LUTs and no UltraRAM for SiTCP-XG.

At 25GbE with a 9000-byte MTU, XTCP delivered 24.70~\gbs{} to the host and
24.08~\gbs{} to the FPGA in separate transfers. The endpoint occupied
23\,385 LUTs, 217 block RAM tiles and 32 UltraRAM blocks. In the sustained
echo test the jumbo-frame prototype averaged 19.51~\gbs{} per direction,
with a fifth percentile of 16.45~\gbs{} over 10~s intervals. For a continuous
data source, the duration of such intervals determines how long upstream
buffering must absorb the difference between input and output rates.

\section*{Data and code availability}
The synthesizable XTCP source, Vivado IP package, RDX051 reference design
and host test tools are available at \url{https://github.com/RFSoC47/XTCP}.
The supplementary material contains individual throughput measurements,
resource counts after routing and time series from the sustained echo tests.

\clearpage
\acknowledgments
The author thanks Y.~Xue and J.~Zhang, technical engineers in the
electrical workshop of the Department of Engineering Physics at Tsinghua
University, for their skilled soldering and patient assistance with board
assembly, without which these measurements would not have been possible.

OpenAI Codex assisted with organizing the bench records, preparing plotting
and verification scripts, and drafting and editing the manuscript.

\bibliographystyle{JHEP}
\bibliography{references}

\providecommand{\href}[2]{#2}\begingroup\raggedright\begin{thebibliography}{10}

\bibitem{yang2024cdex}
H.~Yang, T.~Xue, Q.~Pan, B.~Liang, Y.~Liu and J.~Li, \emph{{Design of a
  triggerless readout electronics system for CDEX-50}},
  \href{https://doi.org/10.1088/1748-0221/19/10/t10005}{\emph{Journal of
  Instrumentation} {\bfseries 19} (2024) T10005}.

\bibitem{li2020}
H.-x.~Li, J.~Zhang, Y.~Ding, S.-s.~Cui, Z.~Wang and X.-s.~Jiang, \emph{{FPGA
  implementation of 10 G Ethernet-based DAQ systems for pixel detectors}},
  \href{https://doi.org/10.1007/s41605-019-0145-8}{\emph{Radiation Detection
  Technology and Methods} {\bfseries 4} (2020) 31}.

\bibitem{uchida2008}
T.~Uchida, \emph{{Hardware-Based TCP Processor for Gigabit Ethernet}},
  \href{https://doi.org/10.1109/tns.2008.920264}{\emph{IEEE Transactions on
  Nuclear Science} {\bfseries 55} (2008) 1631}.

\bibitem{sitcpxg}
{Bee Beans Technologies}, ``{SiTCP-XG Library}.''
  \href{https://www.bbtech.co.jp/en/products/sitcp-xg-license/}{Product
  documentation}.

\bibitem{bauer2014}
G.~Bauer, T.~Bawej, U.~Behrens, J.~Branson, O.~Chaze, S.~Cittolin et~al.,
  \emph{{10 Gbps TCP/IP streams from the FPGA for High Energy Physics}},
  \href{https://doi.org/10.1088/1742-6596/513/1/012042}{\emph{Journal of
  Physics: Conference Series} {\bfseries 513} (2014) 012042}.

\bibitem{sidler2015}
D.~Sidler, G.~Alonso, M.~Blott, K.~Karras, K.~Vissers and R.~Carley,
  \emph{{Scalable 10Gbps TCP/IP Stack Architecture for Reconfigurable
  Hardware}},  in \emph{2015 IEEE 23rd Annual International Symposium on
  Field-Programmable Custom Computing Machines}, pp.~36--43, 2015,
  \href{https://doi.org/10.1109/fccm.2015.12}{DOI}.

\bibitem{ruiz2019}
M.~Ruiz, D.~Sidler, G.~Sutter, G.~Alonso and S.~L{\'o}pez-Buedo, \emph{{Limago:
  An FPGA-Based Open-Source 100 GbE TCP/IP Stack}},  in \emph{2019 29th
  International Conference on Field Programmable Logic and Applications (FPL)},
  pp.~286--292, 2019, \href{https://doi.org/10.1109/fpl.2019.00053}{DOI}.

\bibitem{rfc9293}
{W. Eddy (ed.)}, ``{Transmission Control Protocol (TCP)}.''
  \href{https://www.rfc-editor.org/rfc/rfc9293}{RFC 9293, RFC Editor}, 2022.
\newblock 10.17487/RFC9293.

\bibitem{forencichEthernet}
A.~Forencich, ``{Verilog Ethernet}.''
  \url{https://github.com/alexforencich/verilog-ethernet}.

\bibitem{forencichAxis}
A.~Forencich, ``{Verilog AXI stream components}.''
  \url{https://github.com/alexforencich/verilog-axis}.

\bibitem{amdPG210}
{AMD}, ``{10G/25G High Speed Ethernet Subsystem Product Guide (PG210)}.''
  \href{https://docs.amd.com/r/en-US/pg210-25g-ethernet/Subsystem-Overview}{Subsystem
  overview}, accessed 22 November 2026.

\bibitem{ieee8023}
{IEEE}, ``{IEEE Standard for Ethernet}.''
  \href{https://standards.ieee.org/ieee/802.3/10422/}{IEEE Std 802.3-2022},
  2022.
\newblock 10.1109/IEEESTD.2022.9844436.

\bibitem{rfc6582}
T.~Henderson, S.~Floyd, A.~Gurtov and Y.~Nishida, ``{The NewReno Modification
  to TCP's Fast Recovery Algorithm}.''
  \href{https://www.rfc-editor.org/rfc/rfc6582}{RFC 6582, RFC Editor}, 2012.
\newblock 10.17487/RFC6582.

\bibitem{dg25core}
{Design Gateway}, ``{TOE25G IP Core Data Sheet}.''
  \href{https://www.dgway.com/products/IP/TOE25G-IP/dg_toe25gip_data_sheet_xilinx/dg_toe25gip_data_sheet_xilinx.htm}{Rev.
  1.5, table 1}, 2023.

\bibitem{dg25demo}
{Design Gateway}, ``{TOE25G-IP with CPU Demo Instruction}.''
  \href{https://www.dgway.com/products/IP/TOE25G-IP/dg_toe25gip_cpu_instruction/dg_toe25gip_cpu_instruction.htm}{Rev.
  1.3, figs. 3-4 and 3-8}, 2023.

\end{thebibliography}\endgroup
\end{document}